\input{amssymb.sty}
\input{euscript.sty}
\documentstyle [prl,aps,floats] {revtex}
\input{psfig.sty}
\input{epsf.sty}

\begin{document}
\draft

\twocolumn[\hsize\textwidth\columnwidth\hsize\csname@twocolumnfalse\endcsname 
\title{Coexistence and Criticality in Size-Asymmetric 
Hard-Core Electrolytes}
\author{Jos\'e Manuel Romero-Enrique,\cite{byline1} G. Orkoulas, 
Athanassios Z. Panagiotopoulos,\cite{byline2} and Michael E. Fisher} 
\address{Institute
for Physical Science and Technology, University of Maryland,
College Park, Maryland 20742-2431}
\date{\today}
\maketitle
\begin{abstract}
Liquid-vapor coexistence curves and critical parameters for
hard-core 1:1 electrolyte models with diameter ratios
$\lambda =\sigma_{-}/\sigma _{+}=1$ to 5.7 have been studied by 
fine-discretization Monte Carlo methods. 
Normalizing via the length scale 
$\sigma _{\pm}=\frac{1}{2}(\sigma _{+}+\sigma _{-})$, 
relevant for the low densities in question, both 
$T_{c}^{\ast}\, (=k_{B}T_{c}\sigma _{\pm}/q^{2})$ and 
$\rho _{c}^{\ast }\, (=\rho_{c}\sigma _{\pm}^{3})$ decrease rapidly 
(from $\simeq \! 0.05 $ to 0.03 and 0.08 to 0.04, respectively) 
as $\lambda$ increases. 
These trends, which unequivocally contradict current theories, are closely
mirrored by results for tightly tethered dipolar dimers 
(with $T_{c}^{\ast }$ lower by $\sim 0$-$11\%$ and $\rho _{c}^{\ast}$ 
greater by $37$-$12\%$).
\end{abstract}
]

The formation of coexisting fluid phases of different electrolyte
concentrations in ionic solutions has been the topic of numerous
recent experimental \cite{Kle99}, theoretical \cite{Gon99,Raineri}
and simulation studies \cite{Orkoulas1,cai97,ork99,yan99}. The
simplest models for electrolytes---the ``primitive models''--- 
treat the solvent as a uniform dielectric continuum.
The most studied system is the restricted primitive model
(RPM) that consists of
equisized hard spheres, half carrying a charge $+q$ and
half $-q$. Recent simulations \cite{Orkoulas1,cai97,ork99,yan99} agree
with respect to the critical temperature and density for the 
vapor-liquid transition, finding 
the remarkably low values $T_{c}^{\ast} \simeq 0.05$ and
$\rho_{c}^{\ast} \simeq 0.07$ \cite{ork99}: see (\ref{trho}).
By contrast to the RPM, the effects of charge and size asymmetry have not been 
extensively analyzed either theoretically or via simulation. 

Here we focus on the effects of \emph{size asymmetry} on gas-liquid
coexistence and critical parameters by studying hard-core 
primitive models for 1:1 electrolytes that have \emph{no restrictions} on the
relative magnitude of the diameters of the $+$ and $-$ ions,
 i.e., the so-called size-asymmetric 
primitive model (SAPM)\cite{Raineri}.  
The first claim to treat size asymmetry theoretically appears
already in Debye and H\"uckel's original paper \cite{Debye,Zuckerman} and 
extensions invoking Bjerrum ion-pairing have been analyzed \cite{Zuckerman}.
Other mean potential approaches include the symmetrized
Poisson-Boltzmann and modified Poisson-Boltzmann
\cite{Outhwaite} schemes.    
The mean spherical approximation (MSA)
\cite{Gon99,Raineri,Outhwaite} and hypernetted-chain 
(HNC) \cite{Belloni} integral equations have also been applied.
Currently, however, there are no
simulation results available to check these various theories. 

This work, which extends 
\cite{lrpm}, provides a first study of the effects of
size asymmetry on both critical
parameters and liquid-vapor coexistence \cite{jose}. 
We find, in fact, a systematic trend of $T_c$ and $\rho_c$ with 
increasing size asymmetry that directly conflicts 
with the principal theories cited. 

To be specific, we consider a system of $N$ hard spheres of diameter 
$\sigma_+$ carrying charges $+q$, and $N$ of diameter $\sigma_-$ carrying 
charges $-q$. The interaction energy between two nonoverlapping ions, 
$i$ and $j$, of charges $q_i$ and $q_j$ ($=\pm q$) separated by distance 
$r_{ij}$ is $U_{ij}=q_{i}q_{j}/Dr_{ij}$, where $D$ represents the 
dielectric constant of the solvent which will be set to unity.
The hard-sphere interactions are supposed additive so the ($+$,$-$)-ion 
collision diameter is
$\sigma_{\pm} = \case{1}{2} \left( \sigma_{+} + \sigma_{-} \right)$. 
This, in fact, provides the basic length scale appropriate for 
defining both the reduced temperature and the reduced density via 
\begin{equation}
T^{\ast} = k_{B}TD\sigma_{\pm}/q^{2} \quad \textrm{and} \quad 
\rho^{\ast} = \rho \sigma_{\pm}^{3} \ ,\label{trho}
\end{equation}\cite{Gon99,Orkoulas1} 
where $\rho = 2N/V$ is the \emph{total} 
ionic number density. Other definitions 
of the reduced density are, of course, viable \cite{Raineri,Outhwaite} 
and might be advantageous at high densities. 
However, at the low densities of interest 
here ($2\rho_{c}^{\ast} \lesssim 0.2$) the formation of ion pairs, triples, 
chains and rings (see Fig. \ref{snapshot} below) is controlled almost 
exclusively by $\sigma_{\pm}$, which remains well defined even 
if $\sigma_{+}$ \emph{or} $\sigma_{-}$ vanishes yielding point 
ions \cite{Gillan2}. The reduced simulation box length is defined 
similarly via 
$L^{\ast} = L/\sigma_{\pm}$. 

The key parameter for our study is the ratio of diameters of positive and 
negative ions, namely,
\begin{equation}
\lambda = \sigma_{-} / \sigma_{+} \ . \label{ratio}
\end{equation}
Because of symmetry with respect to the exchange of $+$ and $-$ 
ions, only $\lambda \geq 1$ need be considered.  

In addition to the ionic systems, we have studied, for the sake of 
comparison \cite{Patey1}, tightly tethered dipolar dimer systems 
consisting of 
$N$ pairs of a positive and a negative
ion restricted to remain at separations $\sigma_d$ satisfying
$\sigma_{\pm} \leq \sigma_d \leq 1.02 \sigma_{\pm}$.
Interactions of the tethered dimers are otherwise identical
to those of the ions.

We adopt the methodology of \cite{lrpm}. 
Neutral grand-canonical fine-discretization Monte
Carlo simulations (characterized by a temperature, $T$, and a
chemical potential for a pair of unlike ions, $\mu$) have been
performed on cubic boxes of length $L$, under periodic boundary
conditions. The positions available to each ion are the sites of
a simple cubic lattice of spacing $a$.  A ``lattice refinement''
parameter $\zeta=\sigma_\pm/a$ is introduced, so that when $\zeta
\to \infty$ the continuum is recovered. 
For values of $\zeta \geqslant 3$, the RPM displays a 
gas-liquid coexistence curve that approaches the continuum case quite
closely already for $\zeta=5$ \cite{lrpm}. 
In a Lennard-Jones
fluid studied using $\zeta=10$, the phase envelope and
critical points of the lattice and continuum systems were 
equal within the simulation uncertainties \cite{llj}. 
The structure of
the liquid at short distances, as judged by the pair 
correlations, was also the same.

We have thus adopted a refinement parameter of
$\zeta=10$. For the largest value of $\lambda$ we explore, 
namely $\lambda \simeq 5\case{2}{3}$, the diameter of the smaller sphere is 
$\sigma_{+} = 3a$. 
When $\lambda = 1$, typical correlation
functions for both like and unlike ion pairs then agree  
to within graphical accuracy with the corresponding continuum 
results (for the same $T^{\ast}$, $\rho^{\ast}$ and $L^{\ast}$).
For higher values of
$\lambda \, (> 1)$, the discretization effects on the ($+$,$-$) and
($-$,$-$) correlation functions decrease, while
those on the ($+$, $+$) correlation functions
increase, as $\sigma_{+}/a$ becomes smaller; 
but the latter are
very small at short distances because of the strong
repulsions.

\begin{table} [t]
\caption{Critical parameters, $T_{c}^{*}(L^{*})$ and 
$\rho_{c}^{*}(L^{*})$, for the $\lambda=1$, $\zeta=10$ ionic models 
with two values of $\epsilon_{\infty}$. (The $1\sigma$ statistical 
uncertainties refer to the last decimal place.) }
\vspace{1mm}
\begin{tabular} {lcccc}
$L^*$ &~$100T_{c}^{*}$: $\epsilon_{\infty}=1$, &
$\epsilon_{\infty}=\infty$.&$100\rho_{c}^{*}$: $\epsilon_{\infty}=1$, 
&~$\epsilon_{\infty}=\infty$.~~~\\
\hline 
12 &~~~~~~~~~~5.09(1) &~~~~4.97(1)~~~~~~~~&~~~~~~~7.0(3) & 8.0(2)~~~~~~\\	
15 &~~~~~~~~~~5.03(1) &~~~~4.96(1)~~~~~~~~&~~~~~~~7.0(2) & 8.2(3)~~~~~~\\
18 &~~~~~~~~~~5.00(1) &~~~~4.96(1)~~~~~~~~&~~~~~~~7.4(2) & 7.9(2)~~~~~~\\
\end{tabular}
\label{table1}
\end{table}

The fine-lattice technique embodies precomputation  
and subsequent look-up of the
Coulomb interaction between any two lattice sites,
including all periodic images. 
The Ewald sums were performed with conducting (``tin-foil'') boundary 
conditions, i.e., $\epsilon_\infty=\infty$; but for
$\lambda=$1, vacuum boundary
conditions ($\epsilon_\infty=1$) were also used to allow
comparisons with \cite{ork99}: see Table \ref{table1} and below.

\begin{figure} [b]
\centerline{\psfig{figure=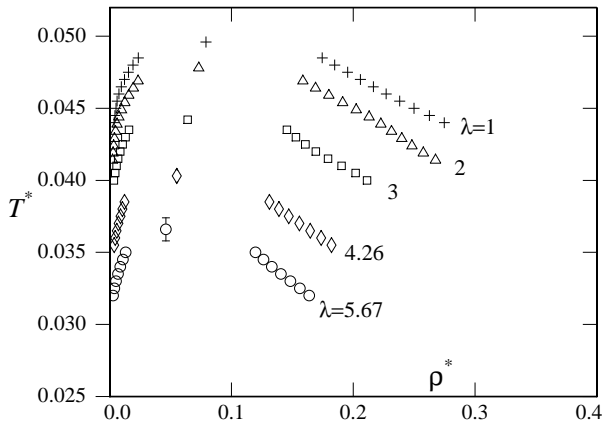,width=8.25cm,angle=90}}
\caption{Phase diagrams for ionic systems with various size-asymmetries 
$\lambda = \sigma_{-}/\sigma_{+}$. 
Statistical uncertainties
are shown only when larger than symbol sizes.}
\label{phase}
\end{figure}

Biased insertions and deletions of pairs of unlike ions were
performed for ionic models, following \cite{Orkoulas1}. 
Our tethered dimers, have 318
distinct configurations on the lattice. Dimers were inserted by 
randomly placing the $-$ ion and selecting one of the 318 positions 
for the $+$ ion. 

Histogram reweighting techniques were used to obtain the
vapor-liquid envelopes up to $T\lesssim 0.98T_c$ \cite{fs88}. 
Effective critical points for given $L^*$ were
estimated using mixed-field finite-size scaling methods
\cite{wilding}, assuming Ising-type criticality.
To discern a systematic dependence on $\lambda$, this approach  
should be satisfactory even though recent results \cite{yy} (which 
indicate that the pressure should also enter the field-mixing) cast 
doubts on its full reliability.

The two Ewald-sum boundary conditions for $\lambda=1$ yield 
different critical values (see Table \ref{table1}) but extrapolation 
to $L^{*}=\infty$ gives 
$T_c^*=0.0495(2)$ for both cases and
$\rho_c^*= 0.078(5)$ for $\epsilon_\infty=1$ and
$0.079(5)$ for
$\epsilon_\infty=\infty$. 
The agreement is excellent; but since the $\epsilon_\infty=\infty$ results
vary less with $L^{\ast}$ they were used for the further
simulations. 

These $\lambda=1$ results should approximate well
those of the continuum RPM: recent studies \cite{cai97,ork99,yan99} 
yield 0.0488(2)-0.0490(3) for $T_c^*$ and 0.062(5)-0.080(5) for $\rho_c^*$. 
Our 1\% larger value of $T_c^*$ may be due to 
lattice discretization. Indeed, previous simulations 
\cite{lrpm}, found that both $T_c^{*}(L^{*})$ and $\rho_c^{*}(L^{*})$ 
estimates decreased slightly as $\zeta$ increased.  
 
\begin{figure} [b]
\centerline{\psfig{figure=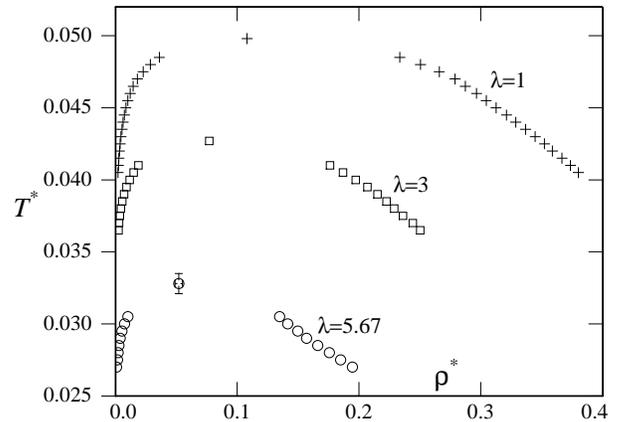,width=8.25cm,angle=90}}
\caption{Phase diagrams for tethered dipolar dimers.} 
\label{phase2}
\end{figure}

The calculated phase diagrams for the ionic and tethered dimer 
systems are shown in
Figs. \ref{phase} and \ref{phase2}, respectively.  
The critical points, which are listed in Table \ref{table2}, 
were calculated \cite{wilding} using $L^{*}=18$ data 
while the subcritical coexistence curves were obtained 
using $L^*=12$. The effects of size asymmetry are clearly strong,
displaying a marked downward shift in $T_{c}^{*}$ and $\rho_{c}^{*}$  
as $\lambda$ increases. 
(The lower values of $T_{c}^{*}$ 
result in smaller
Monte Carlo acceptance ratios and increasing sampling difficulties.) 
 
\begin{table} [t]
\caption {Dependence on $\lambda$ of estimated critical parameters for 
(a) ionic and (b) tethered dimer models.} 
\vspace{1mm}
\begin{tabular} {cccc} 
$~~~~~~\lambda$&$T_c^{*}\times 10^2$&$-\mu_c/k_{B}T_c$
&$\rho_c^{*}\times 10^2$~~~~~~~\\
\hline
(a)~~1&4.96(1)&1.3424(1)&~~7.9(2)~~~~~~~~\\
~~~~~~2&4.79(1)&1.3347(1)&~~7.3(2)~~~~~~~~\\
~~~~~~3&4.42(1)&1.3189(1)&~~6.4(4)~~~~~~~~\\
~~~~~~~~~4.26&4.03(2)&1.3009(4)&~~5.5(3)~~~~~~~~\\
~~~~~~~~~5.67&3.66(8)&1.2880(8)&~~4.6(1)~~~~~~~~\\
\hline
(b)~~1&4.98(1)&1.3210(1)&10.8(3)~~~~~~~~\\
~~~~~~3&4.27(1)&1.3005(1)&~~7.7(2)~~~~~~~~\\
~~~~~~~~~5.67&3.28(7)&1.2795(4)&~~5.2(3)~~~~~~~~\\
\end{tabular}
\label{table2}
\end{table}

\begin{figure} [t]
\centerline{\psfig{figure=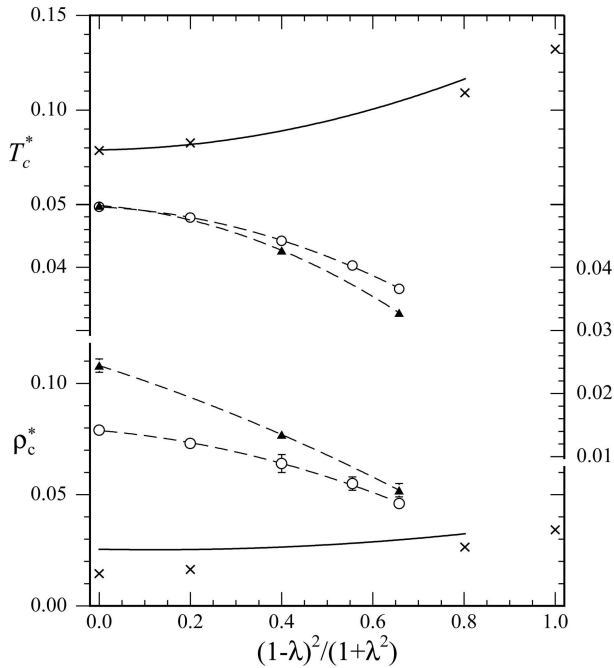,width=8.25cm}}
\caption{Reduced critical temperatures, $T_c^*$, and densities,
$\rho_c^*$, as a function of the asymmetry parameter, 
$\omega(\lambda) = (1-\lambda)^{2}/(1+\lambda^{2})$, as found by 
simulations (i) for ionic systems (open circles) and (ii) for tethered 
dimers (solid triangles); and as predicted by theory using (iii) the 
MSA (energy route) [2] (crosses) and (iv) extended Ebeling-Grigo 
theory (in the EG-Eb approximation) [3] (solid lines).} 
\label{tc}
\end{figure}

For $\lambda=1$, the critical temperatures of the ionic and dimer systems 
seem almost identical. Indeed, although $\rho_c$ is about 37\% 
higher, the overall phase behavior of the dimers is quite similar 
to that of the ionic systems, as stressed by Shelley and Patey \cite{Patey1}. 
When $\lambda$ increases, the critical temperatures of the 
dimers fall more rapidly but the critical densities approach those 
for ions, differing by only 11\% at 
\mbox{$\lambda \simeq 5.67$}: see Fig. \ref{tc} where these results 
are depicted graphically vs the asymmetry parameter 
$\omega(\lambda) \equiv (1-\lambda)^{2}/(1+\lambda^{2})$. 
This parameter respects 
the symmetry under exchange of $+$ and $-$ ions and increases monotonically 
from $\omega(1)=0$ for the RPM, up to $\omega(\infty)=1$, 
for point ions. Extrapolating
our data to the point-ion limit, $\omega=1$, suggests a possibly common 
critical density, $\rho_{c}^{*}$, for ions and dimers of $\sim$0.015, with 
distinct critical temperatures, $T_{c}^{*}$, in the ranges 0.020-0.025 and 
0.012-0.018. To check these speculations, however, may not be easy!

Also shown in Fig. \ref{tc}, as crosses, are the predictions of the MSA 
(using the energy route) for $\lambda=1$, 2, 10 and $\infty$ \cite{Gon99}. 
The absolute differences in $T_{c}^{*}$ and $\rho_{c}^{*}$ were to be 
anticipated (see, e.g., \cite{Fisher}); but it is striking that the predicted 
changes with $\lambda$ in both $T_{c}^{*}$ and $\rho_{c}^{*}$ are 
\emph{opposite} to those revealed by the simulations. The same 
failure to predict 
the correct trends is seen in the various calculations of 
Raineri \emph{et al.} \cite{Raineri} as illustrated by the solid curves in 
Fig. \ref{tc}. These derive from extensions of the Ebeling and Grigo theory 
(which employs Bjerrum ion pairing). (See also \cite{Fisher}.)

The conflict of our data with the theories cited is, perhaps, not so 
surprising when one recognizes the large degree of pair association that 
occurs already in the critical region of the RPM ($\lambda=1$) 
\cite{Fisher,Gillan,Patey2}. Indeed, the presence of many such closely 
coupled dipolar pairs is what motivated the comparison with 
charged dumb-bells \cite{Patey1} and our tethered dimer systems. 
But, as realized for some time \cite{Gillan2,Gillan,Osipov} and evident in 
sample configurations such as that in Fig. \ref{snapshot}, dipolar systems 
undergo significant aggregation, primarily forming ($+$, $-$) chains or 
``living polymers.'' Theories which mainly address the pair correlations 
cannot readily do justice to the geometrical aspects of the formation and 
interaction of such chains \cite{Osipov}.  

Conversely, some insight into the lowering of $T_{c}^{*}$ as $\lambda$ 
increases may be gained by examining how the ground-state binding energies, 
$E_{b}^{*}=E_{b}D\sigma_{\pm}/q^2$, of various specific configurations 
depend on $\lambda$. Thus for a neutral cluster of four ions (or two dipolar 
dimers) we find $E_{b}^{*}=2.586$ for a square ``ring'' when 
$1 \leq \lambda \leq 1+\sqrt{2}$; but $E_{b}^{*}$ falls smoothly when 
$\lambda$ exceeds 2.414 until, at $\lambda \simeq 8.26$, 
the lowest energy 
configuration switches from a planar diamond to a straight chain with 
$E_{b}^{*}=2.333$. Similarly, if two long ($+$, $-$) chains are brought 
together, the excess binding energy per ion is 
$E_{b}^{*}\simeq 0.752$ when $1 \leq \lambda \leq 2.414$ but 
decreases smoothly to 0.698 when $\lambda$ grows larger. 
Other examples exhibit similar effects again rationalizing the 
observed drop in $T_{c}^{*}(\lambda)$.   

\begin{figure} [b]
\centerline{\psfig{figure=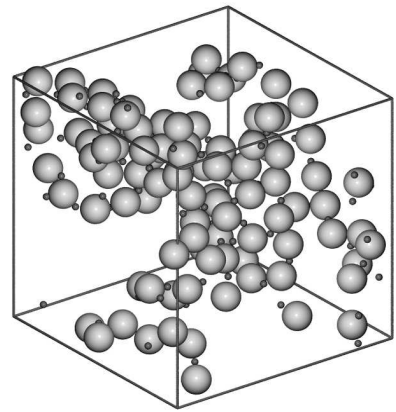,width=8cm}}
\caption{Snapshot of an ionic configuration with 
$\lambda=5\case{2}{3}$ and $L^{*}=15$ at $T^{*}=0.0374 \simeq 1.02T_{c}^{*}$ 
and instantaneous density $\rho^{*}=0.059 \simeq 1.27\rho_{c}^{*}$.}
\label{snapshot}
\end{figure}

The fact that tethered dimers show lower $T_{c}^{*}$ values (for 
$\lambda >1$) than ionic systems, supports the view that \emph{free ions} 
assist in lowering the liquid free energy \cite{Fisher}. The result 
that the dimers require larger densities 
to stabilize the liquid is likewise consistent 
with this idea. In that connection the 
theory of \cite{Fisher}, in which dipolar dimers are solvated by $+$ and 
$-$ free ions (which strongly screen opposite ends of a dimer) may 
reasonably be regarded as approximating the solvation of a dipolar dimer by 
other \emph{dimers}: these will screen by orienting in head-to-tail fashion. 
A better theory is much to be desired but, as 
various attempts illustrate (see, e.g., \cite{gg}), that goal seems elusive. 

\begin{figure} [t]
\centerline{\psfig{figure=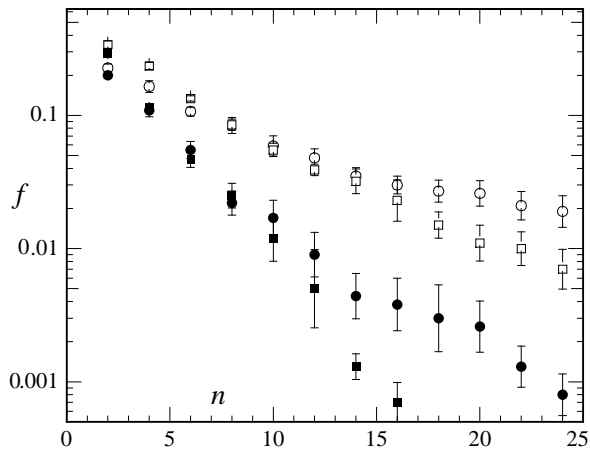,width=8.5cm,angle=90}}
\caption{Fraction, $f$, of ions in neutral clusters of $n$ charged spheres 
in systems of size $L^{*}=18$ at $\rho^{*} \simeq \case{1}{3} \rho_{c}^{*}$. 
Solid symbols denote ionic systems, open symbols tethered dipolar dimers. 
The squares correspond to symmetric ($\lambda=1$) systems at $T^{*}=0.050$, 
the circles to $\lambda=3$ systems at $T^{*}=0.045$.  
From the top downwards (at $n>10$) the reduced densities  
are $\rho^{*}=0.026$, 0.033, 0.019 and 
0.027.}
\label{cluster}
\end{figure}

Finally, in an attempt to quantify some structural differences between 
ionic and tethered dimer systems, cluster densities were sampled for 
selected, comparable conditions: see Fig. \ref{cluster}. 
Gillan's definition of a cluster \cite{Gillan} was used with a clustering 
distance  $R_C^* = R_C/\sigma_\pm = 1.1$. 
It is evident that there are 
more large clusters in the less symmetric systems. 
But tethered dimers have 
much higher fractions of large clusters than do ionic 
systems, even allowing 
for the absence of charged clusters in the former. 

In summary, our fine-discretization simulations 
of hard-core 1:1 electrolyte models have provided unequivocal 
evidence that increasing the size 
asymmetry, measured by the diameter ratio $\lambda=\sigma_{-}/\sigma_{+}$, 
leads to sharp, monotonic drops in appropriately scaled critical 
temperatures and densities [see (\ref{trho})]. Tightly tethered dipolar 
dimers (or dumb-bells \cite{Patey1}) display broadly similar behavior but 
with relatively larger critical densities, and critical 
temperatures that decrease 
faster with increasing $\lambda$. These trends are in severe disagreement 
with current theories \cite{Gon99,Raineri} and present what appear to be 
deep challenges to our theoretical understanding even at a qualitative level. 

We thank G. Stell, J.J. de Pablo, L.F. Rull and G. Jackson  for 
discussions and Professor Stell for providing 
a preprint of his work. 
Funding by the Department of
Energy, Office of Basic Energy Sciences (DE-FG02-98ER14858, AZP)
and the National Science Foundation (Grant No. CHE 99-81772, MEF) is 
gratefully acknowledged. JMRE acknowledges an FPI scholarship and 
Grant PB97-0712 from DGESIC, Spain.


\end{document}